\documentclass[12pt]{article}

\usepackage{newtxtext,newtxmath}
\usepackage{graphicx}
\usepackage{tabularx}
\usepackage{booktabs} 

\usepackage[letterpaper,margin=1in]{geometry}

\renewenvironment{abstract}
	{\quotation}
	{\endquotation}

\date{}

\makeatletter
\renewcommand{\fnum@figure}{\textbf{Figure \thefigure}}
\renewcommand{\fnum@table}{\textbf{Table \thetable}}
\makeatother

\usepackage{scicite}

\usepackage{url}

\def\scititle{
	Four-dimensional video imaging via generative deep learning and a diffuser-encoded image sensor
}
\title{\bfseries \boldmath \scititle}

\author{
	Max~T.~Kauss$^{1}$,
	William~Walker$^{1}$,
	Alexander~Ingold$^{1\dagger}$,
    Jakob~Dammann$^{2}$,\and
    Apratim~Majumder$^{1}$,
    Rajesh~Menon$^{1\ast}$\and
	\small$^{1}$Department of Electrical \& Computer Engineering, University of Utah, Salt Lake City \& 84112, USA.\and
	\small$^{2}$Faculty of Physics, Georg-August-Universität Göttingen, Germany.\and
	\small$^\ast$Corresponding authors: rmenon@eng.utah.edu
}

\begin{document} 

% Insert the title and author list
\maketitle

% Abstract, in bold
% There are strict length limits, and not all formats have abstracts.
% Consult the journal instructions to authors for details.
% Do not cite any references in the abstract.
\begin{abstract} \bfseries \boldmath
Light carries rich information across space, spectrum, polarization, and time, yet conventional cameras capture only a narrow projection of this multidimensional structure. Here we introduce \textbf{4DCam}, a compact imaging platform that records full hyperspectral–polarimetric video in real time. A thin diffuser encodes wavelength-dependent information into single-shot \emph{scatterograms}, captured by a polarization-resolving CMOS sensor that simultaneously measures four linear polarization states. A probabilistic generative network then reconstructs 106 spectral bands spanning 450–850 nm across all polarization channels at 35 frames/s, achieving $<1$\% reconstruction error and providing voxel-wise uncertainty estimates. We use 4DCam to image a live \emph{Betta splendens} fish, uncovering polarization-dependent color modulations that remain invisible to conventional cameras. We experimentally show that the 4D information encoded in the scatterograms markedly improves material discrimination, achieving 96\% accuracy for textile classification and 90\% for camouflage detection, compared with 70\% and 80\%, respectively, using 3D hyperspectral imaging alone. Built entirely from passive optics, 4DCam seamlessly integrates physical encoding, generative decoding, and direct inference, enabling real-time, information-complete optical sensing.
\end{abstract}

% The first paragraph of any Science paper does NOT have a heading
% Nor is it indented
\noindent
% \subsection*{Introduction}
Light encodes far more information than intensity alone. Its spectral composition and polarization state carry signatures of the molecular, structural, and geometric properties of matter, providing a powerful basis for sensing and inference across spatial scales ranging from biological tissues to natural terrains~\cite{tyo2006review, dale2013hyperspectral}. Spectro–polarimetric imaging exploits these dimensions by jointly capturing spectral and polarization signatures, enabling improved material classification, enhanced visibility through scattering media, and structural insight into surfaces and tissues. These capabilities underpin a broad range of applications spanning remote sensing, biomedical diagnostics, environmental monitoring, industrial inspection, and security~\cite{gorbunov2018polarization, Snik2014, lu2014medical, picon2009fuzzy, levenson2006multispectral, aloupogianni2021design, kozun2020multi, shaw2003spectral}.

Yet, realizing this promise in practice remains challenging. High-dimensional imagers typically suffer from optical complexity, low photon throughput, and limited temporal resolution. Conventional hyperspectral and polarization cameras depend on tunable filters, scanning mechanisms, or cascaded dispersive and polarization optics—approaches that are bulky, costly, and highly sensitive to motion and misalignment. Filter-wheel and pushbroom systems, for instance, trade temporal resolution for spectral fidelity, while imaging polarimeters often require multiple sequential frames or separate sensors combined through beamsplitters~\cite{fan2024dispersion,fowler2014compressive,yang2021miniaturization,chen2025computational,yako2023video,wenfull}. These architectures increase system size, optical loss, and susceptibility to error, discarding a substantial fraction of incident light through filtering or beam splitting and thereby limiting sensitivity and dynamic range.

Recent snapshot imaging approaches reduce the need for scanning but introduce new trade-offs. Microfilter mosaic sensors, which generalize color filter arrays to tens of spectral bands or multiple polarization channels, enable single-shot acquisition~\cite{SonyMultispectral,SonyPolarization}. Yet they divide pixel area among channels, lowering per-channel spatial resolution and signal, and they require demanding demosaicking and calibration; only a small fraction of incident light reaches each pixel, compromising low-light performance. Diffractive and metasurface encoders offer an alternative by imprinting spectral \cite{majumder2025high,Bian2024,Hagen2013,lin2023metasurface,zhang2023handheld} or polarization \cite{rubin2019matrix} signatures into a single exposure that are subsequently inverted to recover the desired images. Notably, only one platform has attempted simultaneous spectral and polarization recovery in a single aperture \cite{Zhang2024}: an amorphous-silicon metasurface tiled into large superpixels produces a spectro-polarimetric encoding on a monochrome sensor, which is then decoded to yield a 4D ($x, y, \lambda,$ Stokes) data cube. In practice, this approach exhibits limited spatial resolution (set by the superpixel pitch), low optical throughput (limited by the metasurface efficiency), restricted spectral bandwidth and a constrained field of view, the latter two arising from the angle- and resonance-dependence of the meta-atoms. The experiment therefore employs a telecentric relay to keep chief-ray angles near normal, mitigating angular sensitivity at the cost of increased bulk. More broadly, such nanostructured phase/amplitude optics demand slow, alignment-sensitive nanolithography far more intricate than conventional elements, and they are inherently narrowband and polarization dependent: broadband operation tends to incur substantial efficiency penalties and wavelength-/polarization-dependent losses. Consequently, metasurface imagers have so far achieved either high spectral resolution or full-Stokes polarization coverage—but not both simultaneously with high efficiency, fidelity, and spatial or angular bandwidth. 

\begin{figure}[htb]
\centering
\includegraphics[width=\textwidth]{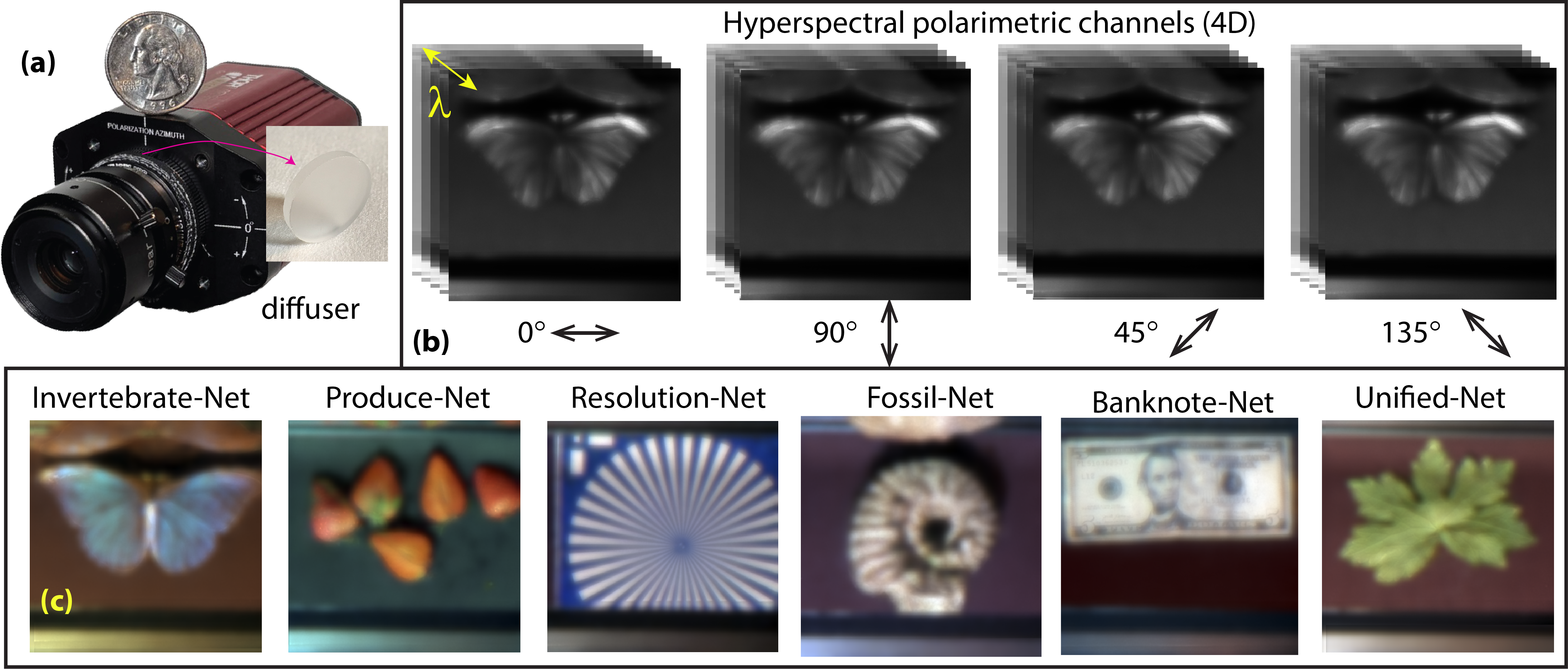}
\caption{\textbf{4DCam for hyperspectral–polarimetric video imaging.}
(a) A polarization-resolving CMOS sensor is augmented with a thin ground-glass diffuser placed in close proximity to the image plane, which encodes wavelength-dependent scatterogram patterns for each polarization channel. (b) Deep neural networks trained for specific application domains reconstruct four-dimensional (4D) hyperspectral–polarimetric datacubes from single grayscale frames. Example from the Invertebrate-net demonstrates simultaneous recovery of four linear polarization states ($0^\circ$, $45^\circ$, $90^\circ$, $135^\circ$) across 106 spectral channels. (c) Exemplary polarization-averaged sRGB renderings of 4D outputs of separate networks trained on distinct object classes—invertebrates, produce, resolution targets, fossils, and banknotes—as well as a unified model trained across all classes are shown.}
\label{fig:Overview}
\end{figure}

Here we introduce the \textbf{4DCam}, a computational spectro–polarimetric imager that combines passive optical encoding with generative deep learning to achieve real-time, 4D video imaging (Fig. \ref{fig:Overview}). A thin ground-glass diffuser encodes wavelength-dependent scatterogram patterns, while a division-of-focal-plane CMOS sensor simultaneously records four linear polarization orientations ($0^\circ$, $45^\circ$, $90^\circ$, $135^\circ$). The resulting single grayscale exposure is decoded by a probabilistic neural network into a 4D datacube, spanning two spatial, one spectral, and one polarization dimension, together with voxel-wise uncertainty estimates. This design eliminates moving parts, dispersive elements, and custom nanofabrication, yielding a compact, alignment-insensitive platform. By simultaneously capturing spectral and polarization information, the imager enhances material discrimination and structural contrast beyond what is possible with either modality alone. We demonstrate hyperspectral reconstruction across 106 spectral bands (450–850 nm) for each of the four polarization channels at 35 frames/s, enabling dynamic 4D visualization and quantitative classification in diverse settings from biological imaging to remote sensing.

Unlike metasurface- or filter-mosaic architectures that sacrifice bandwidth for compactness and suffer from wavelength- and polarization-dependent losses, our approach achieves high throughput and uniform efficiency across all channels using entirely passive, off-the-shelf components. The union of physical encoding and generative inference thus defines a new class of compact, broadband, and uncertainty-aware imagers that extend the dimensionality of light capture for scientific, industrial, and security applications.

\subsection*{Operating principle}
The core principle of 4DCam is the encoding of spectral information into spatial intensity patterns using a thin, passive ground-glass diffuser (Edmund Optics Model 38-784) in front of a polarization-resolving image sensor (Thorlabs Kiralux CS505MUP1, Sony IMX250MZR CMOS sensor, pixel size = 3.45 $\mu$m). A conventional camera lens (f/1.4, focal length = 6 mm, Model Thorlabs MVL6WA) is used such that the field of view is $42^{\circ}$ (horizontal)$\thinspace\times\thinspace36^{\circ}$ (vertical). The diffuser introduces wavelength-sensitive scattering that generates a unique spatial distribution—hereafter referred to as a scatterogram—for each spectral component of the incoming light. Although such diffusers are nominally designed to be spectrally neutral, their random microstructure produces subtle wavelength-dependent intensity variations (see Supplementary Information, Section 1). We harness this chromatic diversity as a spectral encoder, so that each input spectrum yields a distinct, spatially structured scatterogram on the sensor.

At the same time, a micropolarizer array bonded directly to the sensor resolves four linear polarization orientations ($0^\circ$, $45^\circ$, $90^\circ$, and $135^\circ$), producing four de-mosaiced polarization channels from which the Stokes parameters ($S_0$, $S_1$, $S_2$) can be derived. In the presence of the diffuser, each polarization-resolved scatterogram thus encodes both spectral and polarization information, which can then be computationally decoded into hyperspectral–polarimetric images using trained deep neural networks.

At each instant, 4DCam optically compresses a 4D datacube, spatial $(x, y)$, spectral $(\lambda)$, and polarization, into a single 2D sensor frame. The diffuser encodes spectral and spatial features through structured scatterograms, while the micropolarizer array multiplexes polarization across four channels. These raw scatterograms, though unintelligible to the human eye, preserve the complete high-dimensional information of the scene, enabling reconstruction of full 4D hyperspectral–polarimetric video in real time.

Because the optical encoding is highly complex and not readily invertible by analytical methods, as we have previously discussed~\cite{majumder2025high, Wang2014a, Wang2014b, Wang2018MSI}, we adopt a data-driven reconstruction strategy. A probabilistic deep neural network learns the inverse mapping from diffuser-encoded scatterograms to 4D hyperspectral–polarimetric datacubes, producing accurate reconstructions together with per-pixel uncertainty estimates that quantify prediction uncertainty in the absence of ground truth. This learned-inverse approach removes the need for explicit calibration of point-spread functions or computationally intensive iterative solvers, enabling deterministic, real-time reconstructions once trained. Conceptually, the system performs optical compressive sampling, with the neural network acting as a high-dimensional decoder that restores the full spectral and polarization content from a single measurement.

This paradigm extends the broader class of computational imagers that exploit random media or coded optics to multiplex high-dimensional information~\cite{Waller_Diffuser}. In contrast to systems that depend on microfilters, gratings, or custom-fabricated metasurfaces for spectral encoding, our design leverages the natural wavelength-dependent scatterograms of an off-the-shelf ground-glass diffuser and the intrinsic polarization multiplexing of a commercial CMOS sensor. This combination of passive optics and learned inference shifts complexity from hardware to software, yielding a compact, low-cost, and physically robust platform for high-dimensional video imaging.

An overview of the 4DCam architecture and exemplary images are shown in Fig. \ref{fig:Overview}. The system’s versatility is demonstrated across diverse imaging domains, where separate application-specific neural networks are trained for specific datasets using the same physical hardware. Despite differences in training data, ranging from biological to manufactured materials, the reconstructions remain consistent and accurate, underscoring the robustness and generality of the optical encoding. 

To train the reconstruction networks, we sought to learn a direct mapping between the diffuser-encoded scatterograms and their corresponding hyperspectral–polarimetric datacubes. This required paired measurements in which the 4DCam records a polarized scatterogram and a reference hyperspectral camera acquires the ground-truth spectral cube of the same scene. To obtain these data, we implemented a dual-camera acquisition system precisely co-registered to a common field of view (see Supplementary Information, Section 3).

All datasets were acquired on a controlled laboratory stage with the object distance fixed at $\approx80$  cm, under broadband, unpolarized illumination to ensure uniform spectral excitation and minimize ambient variability. Samples were sequentially positioned within the shared field of view of both cameras, enabling synchronized capture of the diffuser-encoded polarization images and the reference hyperspectral cubes. The resulting paired measurements form the basis for supervised learning of the inverse mapping from scatterograms to hyperspectral–polarimetric reconstructions. Additional details of the optical hardware, illumination sources, and calibration protocol are provided in Supplementary Information, Sections 1-3.

\subsection*{Network architecture}
Reconstructing a high-resolution hyperspectral datacube from a single 4DCam polarization channel represents a severely ill-posed, non-linear inverse problem. To retrieve the underlying spectral content, we trained a deep neural network on paired scatterograms and hyperspectral ground-truth images. 

The reconstruction network adopts a U-Net backbone—an encoder–decoder architecture with skip connections that preserves spatial detail while capturing multi-scale context. Unlike conventional deterministic models, our implementation is probabilistic: for each voxel in the reconstructed datacube, the network predicts both the mean intensity and its associated predictive uncertainty, $\sigma$. This is achieved by replacing the standard $\ell_1$ objective with a Laplacian negative log-likelihood loss, enabling the model to jointly minimize reconstruction error and learn calibrated voxel-wise uncertainty. The resulting reconstructions are therefore both quantitative and uncertainty-aware, reflecting the confidence of the model in each predicted voxel.

The total training objective combines complementary criteria enforcing spatial–spectral coherence, spectral fidelity, perceptual realism, and probabilistic consistency:
\begin{equation}
\mathcal{L}_{\text{total}} = \alpha\mathcal{L}_{\text{3DSSIM}} + \beta\mathcal{L}_{\text{SC}} + \gamma\mathcal{L}_{\text{GAN}} + \delta\mathcal{L}_{\text{NLL}},
\label{eq:main_total_prob}
\end{equation}
where $\mathcal{L}_{\text{3DSSIM}}$ promotes structural similarity across the 3D spectral volume, $\mathcal{L}_{\text{SC}}$ maximizes spectral correlation with the reference data, $\mathcal{L}_{\text{GAN}}$ introduces a conditional adversarial constraint to enhance local realism, and $\mathcal{L}_{\text{NLL}}$ enforces probabilistic calibration. The weighting coefficients $(\alpha, \beta, \gamma, \delta)$ were empirically tuned for stable convergence; all hyperparameters and optimization details are provided in Supplementary Information, Section~5.

Each scatterogram produces two outputs: a reconstructed hyperspectral datacube of size $106\times120\times120$ and a corresponding uncertainty map of equal dimensions, for each of the four polarization channels. Together, these form a 4D output of size $212\times120\times120\times4$, yielding interpretable, uncertainty-aware hyperspectral–polarimetric imaging.

To evaluate performance and generalization, the network was trained and tested on five representative datasets—\emph{Banknotes}, \emph{Invertebrates}, \emph{Produce}, \emph{Resolution charts}, and \emph{Fossils/Flora}. Independent models were trained for each polarization channel, and results were averaged across all test subsets (Table~\ref{tab:prob_results}). A sixth unified model, trained on the aggregate dataset, was used to examine cross-domain transfer. Each dataset was split 80:20 into training and validation subsets and augmented sixteenfold through spatial transformations to increase diversity and mitigate overfitting.

\begin{table}[htbp]
\centering
\caption{\textbf{Average reconstruction performance across datasets and polarization states.}
Metrics are averaged over all four dimensions of the test subsets (20\% of each dataset). 
“Test/Total (n)” lists the number of test and total images per dataset.}
\label{tab:prob_results}
\small
\begin{tabularx}{\linewidth}{l *{5}{>{\centering\arraybackslash}X} >{\centering\arraybackslash}X}
\toprule
\textbf{Dataset} & \textbf{SSIM} & \textbf{MSE ($\times 10^{-3}$)} & \textbf{MAE} & \textbf{PSNR} & \textbf{Avg. Uncertainty} & \textbf{Test/Total (n)} \\
\midrule
Banknotes         & 0.956 & 0.355 & 0.00935 & 34.49 & 0.00421 & 14/70 \\
Invertebrates     & 0.955 & 0.480 & 0.00928 & 33.18 & 0.00250 & 18/90 \\
Produce           & 0.951 & 0.602 & 0.00952 & 32.20 & 0.00275 & 40/202 \\
Resolution charts & 0.970 & 0.373 & 0.01077 & 34.28 & 0.00432 & 40/199 \\
Fossils/Flora    & 0.983 & 0.115 & 0.00422 & 39.40 & 0.00427 & 73/366 \\
\midrule
Unified model  & 0.931 & 0.930 & 0.01417 & 30.31 & 0.01001 & 174/871 \\
\bottomrule
\end{tabularx}
\end{table}

Across all datasets, the probabilistic model achieved consistently high structural similarity and low reconstruction error. Dataset-specific models outperformed the unified network, which exhibited a modest decline in accuracy when applied across heterogeneous domains—highlighting both the advantage of domain specialization and the challenge of generalizing a single network across diverse visual categories (Supplementary Fig.~6).

To assess reproducibility and uncertainty calibration, an ensemble of five independently initialized probabilistic networks was trained on the \emph{Invertebrates} dataset, which included butterfly, moth, and beetle specimens from the Natural History Museum of Utah. Figure~\ref{fig:prob_recon_analysis} illustrates representative reconstructions of a preserved \emph{Morpho} butterfly, a neotropical species characterized by strong structural coloration and polarization-dependent reflectance, an ideal benchmark for spectral fidelity. Ensemble reconstructions exhibit excellent spatial and spectral agreement with the reference data (Figs.~\ref{fig:prob_recon_analysis}a–c). Spectra extracted from individual pixels (Figs.~\ref{fig:prob_recon_analysis}d, g) show near-identical profiles among ensemble members and close correspondence with ground truth. The Laplacian probability density functions (Fig.~\ref{fig:prob_recon_analysis}h) reveal narrow, overlapping distributions, indicating consistent uncertainty calibration across models. Together, these analyses confirm that the probabilistic framework delivers stable, reproducible, and uncertainty-aware reconstructions with high spectral fidelity.

A principal strength of this framework lies in its ability to yield per-pixel uncertainty alongside each reconstructed datacube. To evaluate the reliability of these estimates, we analyzed over 29 million pixels across all test images. The predicted uncertainty, $\sigma$ exhibited a strong correlation with the actual reconstruction error (MAE), with a global Pearson coefficient of $r=0.74$. Dataset-level analyses yielded similarly high correlations ($r=0.66$–$0.90$; Supplementary Section~7), confirming that uncertainty maps accurately track local reconstruction quality. The probabilistic U-Net thus provides both high-fidelity hyperspectral reconstructions and calibrated, interpretable uncertainty, an essential capability for autonomous and quantitative imaging systems.

\begin{figure*}[htb]
  \centering
  \includegraphics[width=0.5\textwidth]{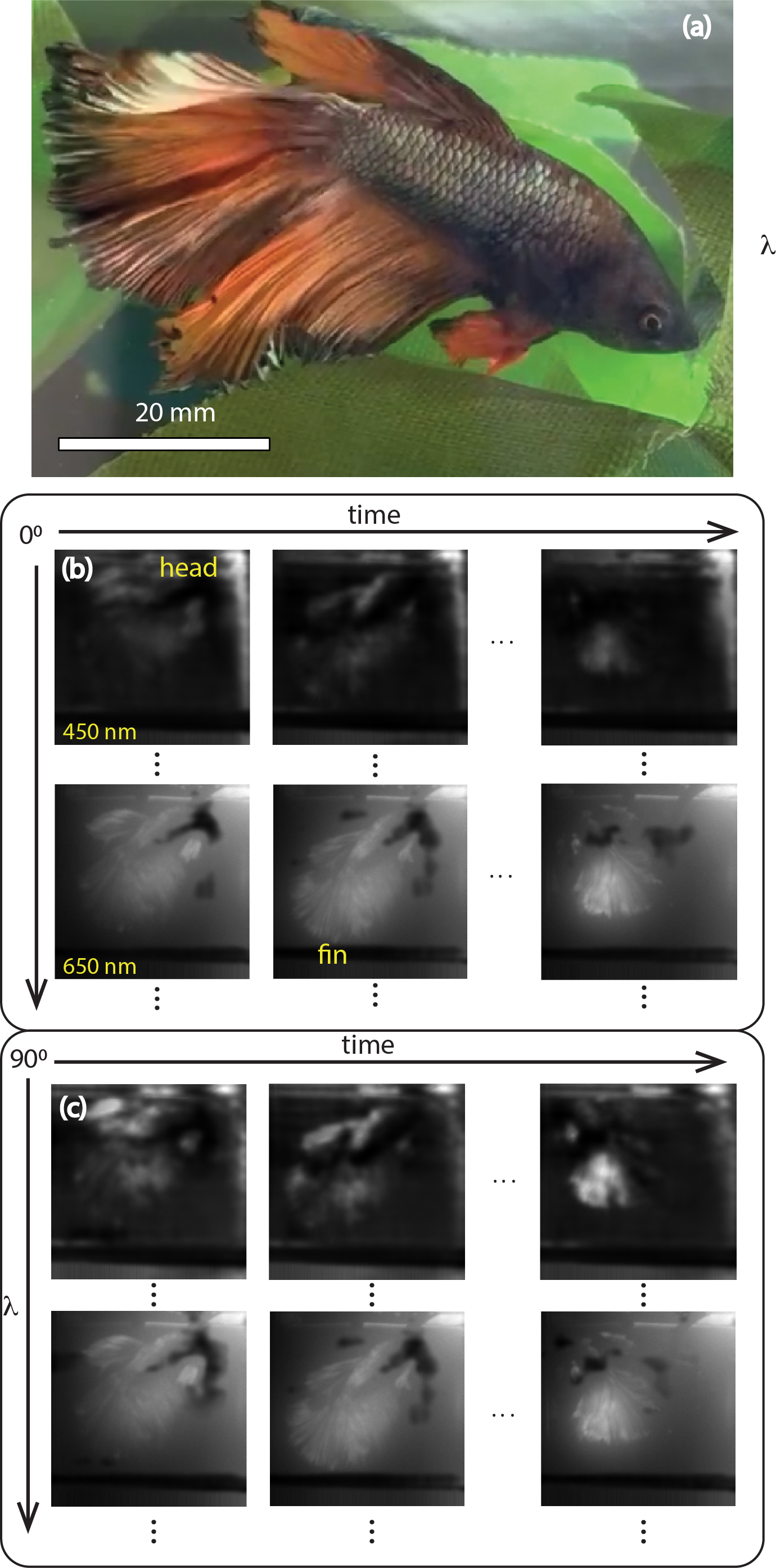}
  \caption{\textbf{4DCam video imaging of a live \emph{Betta Splendens} at 35 frames per second.}
(a) Overview image of the specimen. Representative frames from the reconstructed 4D hyperspectral–polarimetric video, showing spectral ($\lambda$), and polarization (b) $0^{\circ}$ and (c) $90^{\circ} $channels (see Supplementary Videos 1 and 2). Reconstructions were generated using the unified model, which had not encountered similar dynamic, reflective scenes during training. Although no ground-truth data exist at these frame rates, the predicted voxel-wise uncertainty remains comparable to that observed for the training datasets, indicating high reconstruction fidelity and uncertainty in the recovered spectral–polarimetric dynamics.
  \label{fig:video_demo}}
\end{figure*}

\begin{figure*}[htbp]
    \centering
    \includegraphics[width=\textwidth]{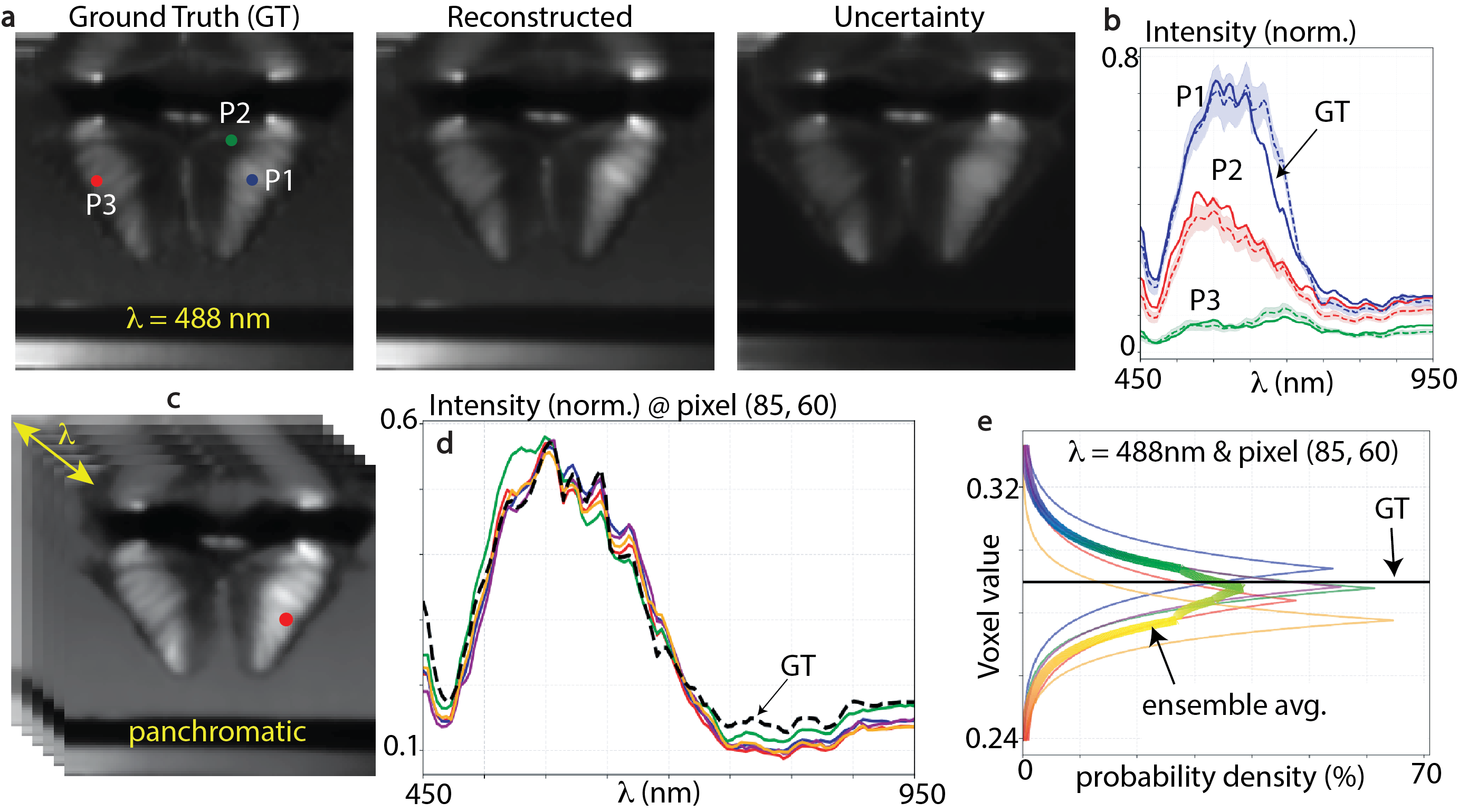}
    \caption{\textbf{Uncertainty-aware hyperspectral reconstructions.} 
    (\textbf{a}) Left to right: Ground-truth (GT), reconstructed, and uncertainty maps at $\lambda=488$~nm for polarization $\theta=0^\circ$. 
    (\textbf{b}) Spectra from labeled points P1–P3 demonstrate close agreement between ground truth (solid lines) and reconstruction (dashed lines), with shaded regions denoting $\pm\sigma$ intervals.
    (\textbf{c}) Full hyperspectral stack and panchromatic image. 
    (\textbf{d}) Spectra from pixel (85,60) reconstructed by five independently trained probabilistic networks closely match both the ground truth and one another.
    (\textbf{e}) Laplacian probability density functions (PDFs) at $\lambda=488$~nm quantify ensemble variability, showing narrow, overlapping distributions indicative of calibrated uncertainty and consistent predictive behavior.
    \label{fig:prob_recon_analysis}}
\end{figure*}

To recover fine spatial detail absent from the supervised hyperspectral training data, optionally we apply a deterministic post-processing step termed smooth multiplicative fusion (see Supplement Section S6). This approach transfers high-resolution spatial structure from the broadband diffuser measurement to each reconstructed spectral band while preserving the spectrally resolved content predicted by the network. The method redistributes measured broadband intensity according to locally estimated spectral proportions, yielding a spatially super-resolved hyperspectral datacube that remains quantitatively consistent with the original reconstruction when averaged back to the native resolution. We note that this smooth multiplicative fusion has only been applied to the images in Figures~\ref{fig:Overview} and ~\ref{fig:video_demo} to demonstrate efficacy.

\subsection*{Hyperspectral Polarimetric Video} 
To demonstrate dynamic 4D imaging, 4DCam was used to record a live \emph{Betta Splendens} (Siamese fighting fish) at 35 frames/s. Each polarization-resolved scatterogram was reconstructed into a hyperspectral–polarimetric video sequence, revealing both spectral contrast and polarization dynamics in real time. Reconstructions were generated using the unified network trained jointly across all datasets. Although the network had never encountered scenes resembling the iridescent, polarization-dependent reflections of fish scales, it produced temporally stable, high-fidelity reconstructions throughout the sequence.

Because conventional hyperspectral cameras cannot operate at comparable frame rates, no ground-truth video exists for validation; instead, the model’s predicted uncertainty ($\sigma$) provides a quantitative measure of reconstruction reliability. Across all frames and polarization angles, the mean uncertainty was $\sigma\approx0.012$—only slightly higher than the testing uncertainty of $\sigma=0.010$—indicating robust performance despite motion blur and specular reflections. Elevated $\sigma$ values were localized primarily to the reflective fins and scales, while the surrounding water remained uniformly low in uncertainty. Together, these results highlight 4DCam’s ability to recover rich spectral–polarimetric information from rapidly evolving natural scenes, demonstrating its potential for real-time 4D biological video imaging.

Figure \ref{fig:video_demo} shows representative frames from a 4DCam video sequence of a moving {\it Betta splendens}, displaying total intensity alongside polarization-resolved spectral responses. Although the system records four linear polarization channels ($0^{\circ}$, $45^{\circ}$, $90^{\circ}$ and $135^{\circ}$; Supplementary Videos 1 and 2), only two orthogonal polarization states are shown here for clarity. The frames reveal pronounced wavelength- and polarization-dependent variations in the reflected light from the fish’s highly iridescent fins. Dark banding patterns emerge from polarization-selective reflections and birefringence within the layered iridophore structures of the scales and fins, which locally alter the polarization state of scattered light. The spatial contrast and distribution of these features evolve with wavelength, indicating dispersive behavior associated with structural coloration. These observations illustrate how polarization-resolved hyperspectral video captures optical anisotropies and wavelength-dependent scattering mechanisms that are not accessible through intensity-only imaging.

This experiment further underscores that the achievable frame rate of hyperspectral–polarimetric imaging is limited only by the native acquisition speed of the sensor. Conventional hyperspectral and polarization cameras rely on sequential spectral or polarization scanning, rendering them too slow to capture rapidly evolving phenomena. In contrast, 4DCam records all spectral and polarization channels simultaneously in a single exposure, enabling dynamic 4D video of transient biological behavior. With task-specific training data, the same approach could be extended to applications that demand rapid, high-dimensional scene understanding, ranging from environmental monitoring and biomedical diagnostics to autonomous sensing, where real-time decisions benefit from simultaneous spectral and polarization awareness.

\subsection*{Information super-resolution via 4D imaging}
\begin{figure}[htbp]
    \centering
    \includegraphics[width=\linewidth]{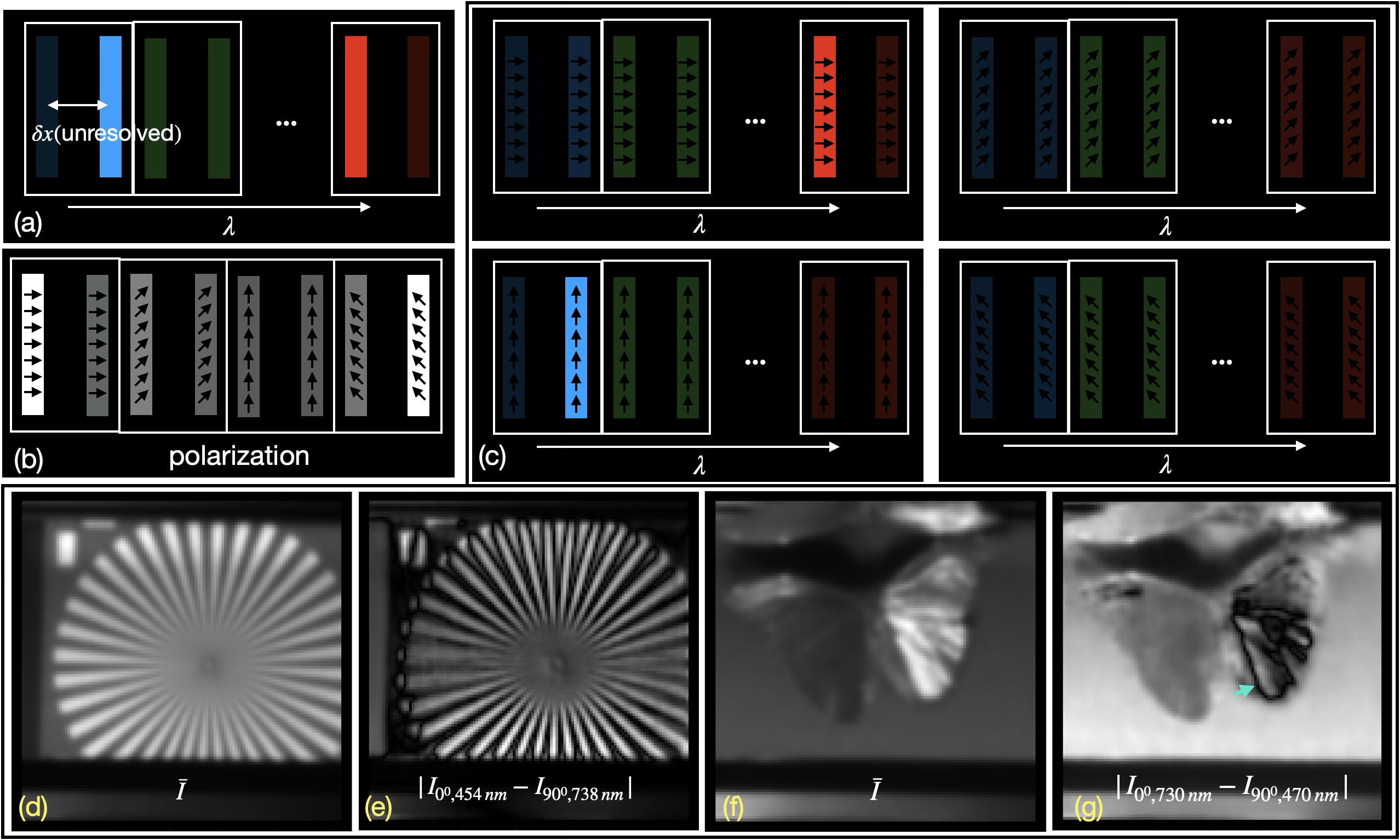}
    \caption{\textbf{Spectro-polarimetric information enhances effective spatial resolution.} 
 (a–c) Illustration showing that two spatially unresolved features, indistinguishable in conventional intensity images, become separable when analysed across additional optical dimensions: (a) through differences in spectral channels, (b) through polarization contrast, or (c) through combined spectro–polarimetric signatures. Harnessing this higher-dimensional information enables effective separation of sub-resolution features. Experimental demonstration using a radial resolution chart. The panchromatic, polarization-averaged baseline image (d) provides limited contrast, whereas the spectro-polarimetric difference image, $|I_{0^{\circ},\lambda=454\,\text{nm}} - I_{90^{\circ},\lambda=738\,\text{nm}}|$, (e) reveals substantially finer structure. Line pairs spaced by 4 sensor pixels (corresponding to an angular resolution of $\approx9$ mrad) exhibit an increase in contrast from 4\% to 24\%, demonstrating resolution enhancement through spectro-polarimetric information (see Supplement Section 9). Validation on a natural specimen, a butterfly wing, further highlights this effect. The baseline panchromatic polarization-averaged image (f) shows low contrast, while the spectro-polarimetric difference image, $|I_{0^{\circ},\lambda=730\,\text{nm}} - I_{90^{\circ},\lambda=470\,\text{nm}}|$, (g) sharply delineates microstructural boundaries, including a narrow dark line (blue arrow) about 1 sensor pixel wide, revealing features beyond the native spatial resolution of the imager.}
    \label{fig:resolution}
\end{figure}

Super-resolution conventionally denotes the capacity to distinguish spatial features separated by distances smaller than the resolution limit by analysing the light they emit or scatter.\cite{aflalo2024optical} Yet, resolution need not be confined to brightness alone. Spatially overlapping or unresolved features—such as two lines separated by a small distance, $\delta x$ (Fig.~\ref{fig:resolution}a)—may nonetheless be distinguished through differences in their spectral (Fig.~\ref{fig:resolution}a), polarization (Fig.~\ref{fig:resolution}b), or combined spectro–polarimetric signatures (Fig.~\ref{fig:resolution}c). When these additional optical degrees of freedom are disentangled, features that are indistinguishable in space become separable in information space. We refer to this broader paradigm as \emph{information super-resolution}: the ability to resolve and classify signals beyond spatial limits by leveraging diversity across wavelength, polarization, and related optical channels.

This principle echoes the logic underlying temporal super-resolution approaches such as localization microscopy,\cite{lelek2021single, basak2025super} where emitters are separated in time rather than space. Extending such strategies to higher-dimensional optical measurements has been challenging, because conventional imaging systems are unable to jointly acquire rich spatial, spectral, and polarization information. Here, we show that resolving features simultaneously across these domains makes it possible to transcend the baseline resolution limit, establishing a regime of high-dimensional information super-resolution that enables discrimination of structures that are otherwise inseparable in space alone.

To highlight this capability, we imaged standard resolution test targets and reconstructed their complete spectro-polarimetric representations. Figure \ref{fig:resolution}d presents the panchromatic, polarization-averaged baseline image of a radial resolution chart. In contrast, Fig. \ref{fig:resolution}e displays a representative spectro-polarimetric difference image, computed as $|I_{0^{\circ},\lambda=454\,\text{nm}} - I_{90^{\circ},\lambda=738\,\text{nm}}|$, which exhibits markedly enhanced contrast. In the baseline image, a line spacing of four sensor pixels yields only 4\% contrast, whereas the spectro-polarimetric difference image achieves $24$\% (see Supplement Section 9). Given a pixel size of 1.85 mm in object space and an object distance of 80 cm, this corresponds to an angular resolution of $\approx9$ mrad. This substantial improvement underscores the ability of high-dimensional imaging to surpass conventional spatial limits: by leveraging the intrinsic diversity among spectro-polarimetric (4D) channels, features that are indistinguishable in conventional intensity images become resolvable.  

To further demonstrate information-based super-resolution in a natural specimen, we imaged a butterfly wing—a sample rich in subwavelength structural color and polarization contrast. The baseline panchromatic, polarization-averaged image is shown in Fig.~\ref{fig:resolution}f. In comparison, one exemplary spectro-polarimetric difference image, computed as $|I_{0^{\circ},\lambda=730\,\text{nm}} - I_{90^{\circ},\lambda=470\,\text{nm}}|$ and shown in Fig.~\ref{fig:resolution}g, reveals a strikingly sharper delineation of the iridescent scales. A narrow dark line, approximately one sensor pixel wide (highlighted by the blue arrow), cleanly separates distinct microstructural domains—features that are not as clearly visible in the baseline image. This pronounced enhancement underscores the ability of spectro-polarimetric contrast to recover spatial detail well beyond the nominal resolution of the imaging system.

\subsection*{Classification of fabrics using 4D information}
\begin{figure}[htbp]
  \centering
  \includegraphics[width=\linewidth]{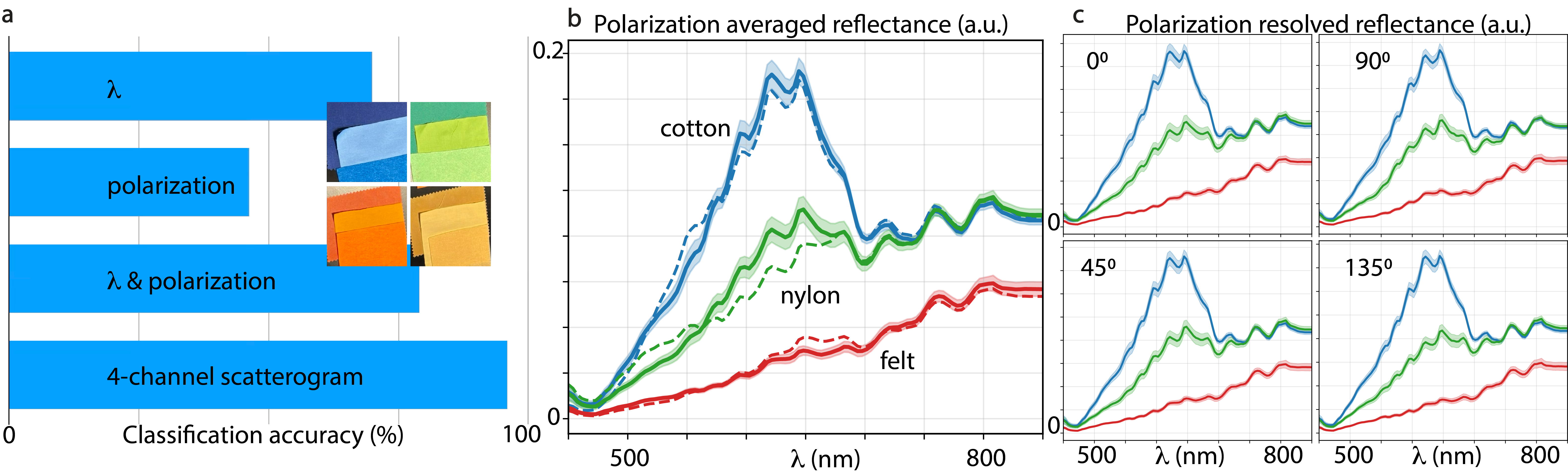}
  \caption{\textbf{Classification of fabrics using 4DCam data.} (a) Average classification accuracy for networks trained on spectral, polarimetric, and fused spectro–polarimetric inputs, as well as on raw scatterograms. All results except the latter are derived from reconstructed hyperspectral–polarimetric images. The uncropped scatterograms were trained on the MobileNetV3 classifier (see Supplement Section 12). (b) Mean reflectance spectra from regions of interest for cotton, felt, and nylon. Dashed lines denote ground truth, solid lines indicate mean reconstructed spectra, and shaded regions represent $\pm1 \sigma$ uncertainty. (c) Representative polarization-resolved spectra from the same ROI example shown in (b), illustrating slight spectral variation across polarization states. Together, these results demonstrate that integrating spectral and polarization cues enhances discriminative power for material classification.
} \label{fig:textile-roi-ResNet}
\end{figure}

We next examined how combining spatial, spectral, and polarization cues, collectively forming a 4D representation, enhances material discrimination. As a representative example, we selected three visually similar textiles: cotton, felt, and nylon (Fig.~\ref{fig:textile-roi-ResNet}a, inset). These fabrics appear nearly identical in color and texture under white light, and exhibit minimal contrast under any single polarization state. While hyperspectral imaging captures material-dependent reflectance spectra, polarization encodes complementary information about surface microstructure and fibre orientation. Integrating these channels therefore offers a route to higher discriminability beyond conventional spectral or polarimetric imaging alone.

A dataset of 162 paired measurements was acquired, each comprising a hyperspectral image cube and a corresponding polarized scatterogram of one of three fabrics (see Supplement Section 10). Textile reconstruction networks were first trained on these paired images using five-fold cross-validation, and the reconstructed hyperspectral outputs—together with the scatterograms—served as inputs for subsequent classifier training. The classification networks were likewise trained and evaluated under five-fold cross-validation to maximize data utilization and to prevent overfitting. As summarized in Fig.~\ref{fig:textile-roi-ResNet}a, classification accuracy increased from 70\% using spectral data alone to nearly 80\% when spectral and polarization information were jointly exploited, compared with 46\% for polarization alone.

To ensure a fair comparison across modalities, features were extracted from the same spatial region and used to train three identical convolutional neural networks. The fused 4D spectro–polarimetric representation yielded the highest accuracy and F1 scores, confirming that polarization encodes microstructural cues that complement spectral signatures. Mean polarization-resolved spectra (Supplement Section 10) reveal substantial overlap among fabrics, highlighting the intrinsic challenge of separating them using a single dimension. Their joint spectro–polarimetric signatures, however, enable robust class separation, establishing 4D imaging as a powerful approach for material identification.

Finally, we tested a lightweight MobileNetV3 classifier \cite{howard2019searchingmobilenetv3} trained directly on the four-channel scatterograms, without explicit image reconstruction (Supplement Section 12). Remarkably, this end-to-end approach achieved 96\% accuracy, demonstrating that direct inference on raw 4D measurements can further streamline analysis while reducing computational cost.

\subsection*{Camouflage Detection}
We next applied 4DCam to the problem of distinguishing natural vegetation from man-made camouflage, an optical classification task that has long challenged conventional imaging approaches. The dataset comprised 100 paired measurements (50 vegetation, 50 camouflage), each containing a hyperspectral image cube and a corresponding polarized scatterogram. Models were trained and validated using five-fold cross-validation to ensure unbiased reconstruction and classification (see Supplementary Information, Section 11).

\begin{figure}[htbp]
    \centering
    \includegraphics[width=\linewidth]{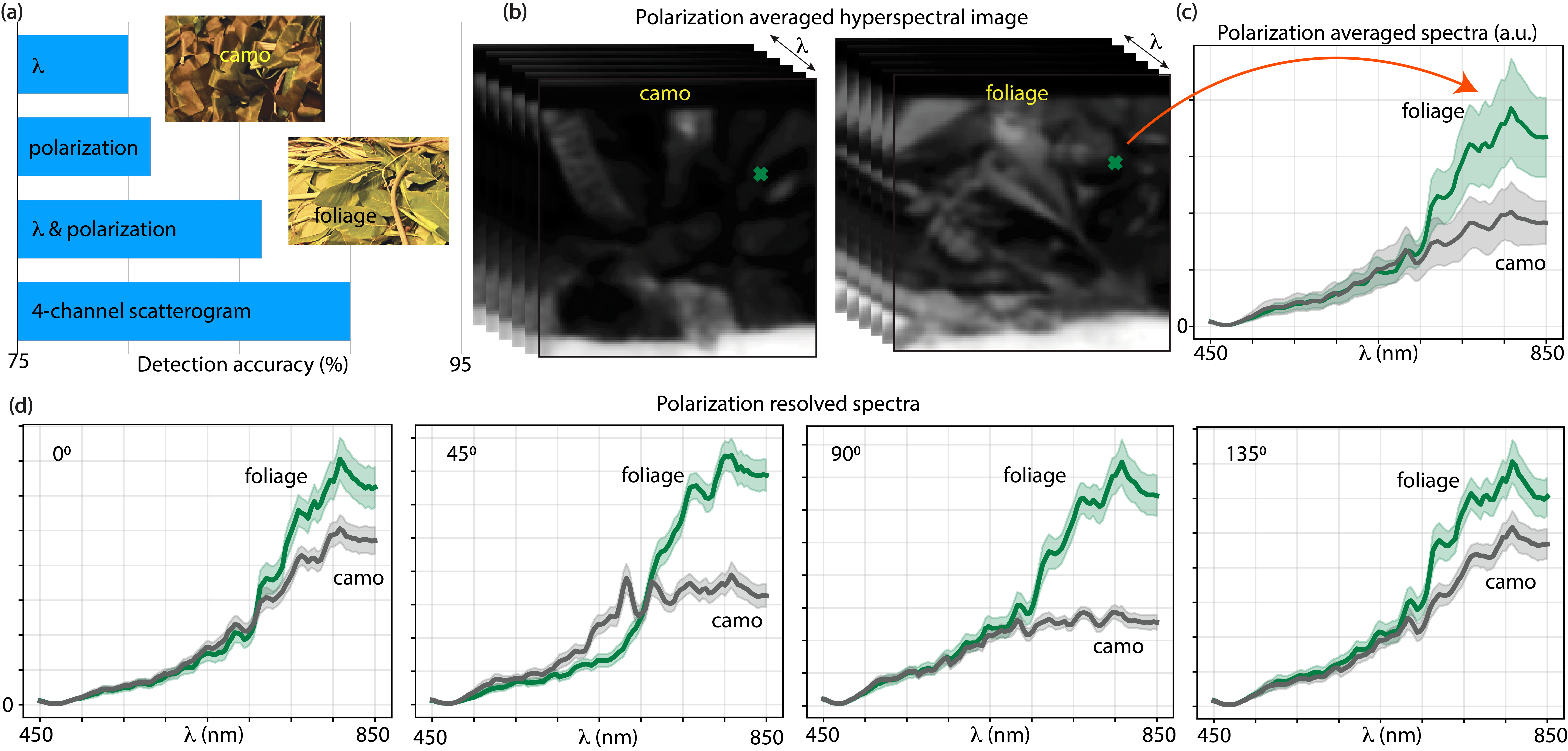}
    \caption{\textbf{Camouflage detection with 4DCam.} 
(a) Classification accuracy improves to $\approx 86 $\% when using 4D hyperspectral–polarimetric reconstructions, compared with $\approx 80$\% when using spectral or polarization cues alone. In the latter cases, data were averaged along the orthogonal dimension. Direct classification of the raw four-channel scatterograms using a lightweight MobileNetV3 network achieved an even higher accuracy of 90\%, notably without requiring any image reconstruction. Inset photographs show example scenes. 
(b) Polarization-averaged hyperspectral images for foliage and camouflage scenes reveal distinct spectral signatures.
(c) Spectra extracted from labeled points in each scene highlight the characteristic near-infrared “red-edge” rise in vegetation reflectance compared with camouflage fabric.
(d) Polarization-resolved spectra from the same points show a markedly stronger and more stable polarimetric signal in foliage, accounting for the improved discrimination achieved with 4D spectro–polarimetric inputs.}
    \label{fig:camo-fig}
\end{figure}

A convolutional classifier was trained on three matched input modalities: spectral-only, polarization-only, and combined spectro–polarimetric (4D) data. As shown in Fig.~\ref{fig:camo-fig}a, the 4D representation achieved the highest accuracy (86\%), outperforming either spectral or polarization inputs (about 80\%). This gain reflects the complementary nature of wavelength- and polarization-dependent cues in separating biological from synthetic materials.

Spectral features captured biochemical contrasts, such as the vegetation “red edge” near 700–850 nm, while polarization revealed microstructural differences arising from surface roughness and birefringence (Figs.~\ref{fig:camo-fig}b–d). Together, these dimensions define a multidimensional contrast space that integrates spectral and structural information, enabling robust, physics-based discrimination between foliage and camouflage.

Finally, when the raw four-channel scatterograms were directly classified using a lightweight MobileNetV3 network, accuracy further increased to 90\% (Supplement Section 12). This reconstruction-free approach, consistent with the textile results, highlights the promise of direct 4D inference for compact, low-power, and high-speed optical sensing.

\subsection*{Conclusion}
4DCam transforms multidimensional imaging from static acquisition to real-time 4D hyperspectral–polarimetric video, capturing the full spatial, spectral, polarization and temporal evolution of light. The generative reconstruction network achieves $<1$\% mean error and 4D-SSIM $\leq$ 0.95 across 106 spectral bands (450–850 nm) and four polarization channels at 35 frames/s, providing voxel-wise uncertainty estimates that quantify reliability in the absence of ground truth. Dynamic imaging of a live \emph{Betta Splendens} reveals time-varying spectral and polarization responses arising from iridescent scales—phenomena inaccessible to conventional hyperspectral systems—demonstrating high-fidelity 4D biological video capture.

Beyond reconstruction, 4DCam delivers measurable performance gains: information super-resolution enhances contrast from 4\% to 24\% for four-pixel line spacing, while 4D material classification improves accuracy from 70\% (spectral-only) and 46\% (polarization-only) to 86\% when both dimensions are fused. A lightweight MobileNetV3 applied directly to raw scatterograms achieves 96\% (textiles) and 90\% (camouflage) accuracy, enabling reconstruction-free, low-power inference.

Crucially, this work establishes a new computational imaging paradigm in which the optical hardware remains fixed and compact, yet application-specific neural networks can be retrained to optimize for distinct sensing tasks or inference objectives, ranging from biological dynamics to remote or autonomous sensing. 4DCam thus represents a SWaP-efficient, uncertainty-aware and reconfigurable imaging architecture that unites physical encoding, probabilistic decoding and task-adaptive inference, advancing toward information-complete cameras capable of learning and reasoning directly from light in real time.

%%%%%%%%%%%%%%%% MAIN TEXT FIGURES %%%%%%%%%%%%%%%
\newpage

\newpage

\newpage

\newpage

\newpage

\newpage

%%%%%%%%%%%%%%%% REFERENCES %%%%%%%%%%%%%%%
\clearpage 
\bibliography{science_template} % for a file named science_template.bib
\bibliographystyle{sciencemag}

%%%%%%%%%%%%%%%% ACKNOWLEDGEMENTS %%%%%%%%%%%%%%%
\newpage
\section*{Acknowledgments}
We thank Syed Noor Qadri for the ground-truth hyperspectral camera. We gratefully acknowledge Alyson Wilkins and Christy Bills from the Natural History Museum of Utah (NHMU) for their assistance in providing and handling invertebrate specimens.  
\paragraph*{Funding:}
Office of Naval Research (ONR) \#N000142512122 and \#N000142212014. 
National Science Foundation (NSF) \#2229036. 
JD acknowledges support from Deutscher Akademischer Austauschdienst German Academic Exchange Service (DAAD RISE).
\paragraph*{Author contributions:\\}
M.T.K. designed and built the camera, performed experiments, developed and trained the machine-learning models, and prepared the manuscript. W.W. conducted experiments, contributed to model training and testing, and assisted with manuscript preparation. A.I. developed the uncertainty-aware machine-learning framework, implemented classification and detection networks, and contributed to writing. J.D. assisted in camera construction, performed experiments, and evaluated early versions of the learning algorithms. A.M. assisted in camera development and data acquisition. R.M. conceived and supervised the project, secured funding, analyzed results, and wrote the manuscript.
\paragraph*{Competing interests:}
There are no competing interests to declare.
\paragraph*{Data and materials availability:}
All data needed to evaluate the conclusions in the paper are present in the paper and/or the Supplementary Materials. 
\paragraph*{Declarations:}
All banknote images used in this study were acquired from genuine currency for non-commercial, scientific analysis, and are presented in compliance with applicable U.S. and international reproduction guidelines to prevent any use as legal tender.

%%%%%%%%%%%%%%%% END OF MAIN TEXT %%%%%%%%%%%%%%%

\newpage

%%%%%%%%%%%%%%%% START OF SUPPLEMENT %%%%%%%%%%%%%%%

% Figures, tables, equations and pages in the supplement are numbered S1, S2 etc.
\renewcommand{\thefigure}{S\arabic{figure}}
\renewcommand{\thetable}{S\arabic{table}}
\renewcommand{\theequation}{S\arabic{equation}}
\renewcommand{\thepage}{S\arabic{page}}
\setcounter{figure}{0}
\setcounter{table}{0}
\setcounter{equation}{0}
\setcounter{page}{1} % not 0 as \newpage already started a supplementary page
% References continue the numbering from the main text.
%%%%%%%%%%%%%%%% MATERIALS AND METHODS %%%%%%%%%%%%%%%

\end{document}